\documentclass[twocolumn,amsmath,amssymb]{revtex4}
\usepackage{graphicx}
\usepackage{color}
\usepackage{float}
\usepackage{hhline}
\usepackage{bm}
\usepackage{pifont}
\usepackage{soul}
\newcommand{\be}{\begin{equation}}
\newcommand{\ee}{\end{equation}}

\begin{document}

\title{A novel coarsening mechanism of droplets \\ in immiscible fluid mixtures}

\author{Ryotaro Shimizu and Hajime Tanaka \footnote{E-mail:tanaka@iis.u-tokyo.ac.jp}}

\affiliation{Institute of Industrial Science, University of Tokyo, 4-6-1 Komaba, Meguro-ku, Tokyo 153-8505, Japan}

\date{Received October 3, 2014}

\begin{abstract}
{\bf 
In our daily lives, after shaking a salad dressing, we see the coarsening of oil droplets suspended in vinegar. Such a demixing process is observed everywhere in nature and also of technological importance. For a case of high droplet density, domain coarsening proceeds with interdroplet collisions and the resulting coalescence. This phenomenon has been explained primarily by the so-called Brownian coagulation mechanism: stochastic thermal forces exerted by molecules induce random motion of individual droplets, causing accidental collisions and subsequent interface-tension driven coalescence. Contrary to this, we demonstrate that the droplet motion is not random, but hydrodynamically driven by the composition Marangoni force due to an interfacial tension gradient produced in each droplet as a consequence of composition correlation among droplets. This alters our physical understanding of droplet coarsening in immiscible liquid mixtures on a fundamental level. } 
\end{abstract}

\maketitle

%%%%%%%%%%%%%%%%%%%%%%%%%%%%%%%%%%%%%%%%%%%%
%% MAINMATTER
%%%%%%%%%%%%%%%%%%%%%%%%%%%%%%%%%%%%%%%%%%%%

Phase separation is one of the most fundamental physical phenomena that lead to pattern formation in various types of systems \cite{gunton1983,binder1987theory,bray1994theory,onuki,puri2009kinetics,cates2012complex}, 
including all kinds of classical condensed matter, biological systems (both ordinary \cite{Membrane} and active matter \cite{Active}), quantum systems such as liquid $^3$He-$^4$He mixtures \cite{Hemixture} and Bose-Einstein condensates \cite{BEC,BEC2}, nuclear matter \cite{Nuclear}, and cosmology \cite{Universe}.  
In many of these systems, momentum conservation plays a crucial role in phase separation.  
The simplest among such systems is a classical binary liquid mixture, where material can be transported by both diffusion and hydrodynamic flow. 
Thus, the relevant field variables are the composition of one of the components 
$\phi$ and the fluid flow velocity $\mathbf{v}$. 
The process of phase separation is then described by the Ginzburg-Landau-type $\phi^4$ free energy and a set of coupled dynamical equations of these two variables satisfying the composition and momentum conservation under the incompressiblity condition $\mathbf{\nabla} \cdot \mathbf{v}=0$ (see Methods for the explicit forms of the equations). 
This model is widely known as model H 
in the Hohenberg-Halperin classification of dynamical critical phenomena and phase separation kinetics
to dynamic universality classes \cite{hohenberg1976}.  
Phase-separation behaviour in many of the above-mentioned systems can basically be described in a similar framework. 
Thus, revealing the fundamental physical mechanism of domain coarsening in model H 
is of significant importance for the general understanding of spinodal decomposition in these systems. 

From a technological point of view, since phase separation starts at nanoscale and its characteristic size proceeds to grow indefinitely with time, we can control the domain size of materials intentionally by freezing the process at a certain time by using liquid-solid transitions such as glass transition and crystallization, by chemical reaction, or by emulsifying liquid droplets with  surfactants.   This strategy is widely used in industrial processes of various materials such as liquid mixtures, emulsions, polymer blends, metallic alloys, ceramics, cosmetics, and foods.  Among various types of phase separation, liquid phase separation is quite important in materials science since a liquid state is most suitable for materials processing. Thus, the understanding of domain coarsening dynamics in the demixing process of  liquid mixtures is not only of fundamental importance but also of crucial technological importance for controlling domain structures of materials. 

The theoretical framework of model H has a very firm basis since it replies solely on the composition and momentum conservation, and there is little doubt on its validity. This phase-field model is widely used for numerical simulations not only in physics, but also in materials science and chemical engineering communities.   
However, the analytical theoretical  analysis of model H is not an easy task due to its non-locality, non-linearity, complex non-local dynamical coupling between $\phi$ and $\mathbf{v}$, and stochastic nature due to thermal noise. 
Nevertheless, the concept of dynamical scaling provides us with a powerful scaling argument for the self-similar domain growth 
in the late stage of phase separation after the formation of a sharp domain interface \cite{gunton1983,binder1987theory,bray1994theory,onuki,furukawa1985dynamic}, 
on the basis of clear physical pictures on the elementary process of domain coarsening. 
Thus, the basic mechanisms of domain coarsening in immiscible liquid mixtures 
are now believed to be reasonably understood  \cite{gunton1983,binder1987theory,bray1994theory,onuki,puri2009kinetics,cates2012complex}. 
Recently, research interests have shifted towards other aspects of phase separation including aging dynamics during the coarsening process \cite{ahmad2012aging}, geometrical features \cite{sicilia2009geometry}, and response functions of coarsening systems \cite{berthier1999response}, inertia effects \cite{furukawa1985dynamic,kendon2001inertial}, and viscoelastic 
effects \cite{onuki,tanaka2000viscoelastic}. 

\begin{figure}[h!]
\begin{center}
   \includegraphics[width =7.0cm]{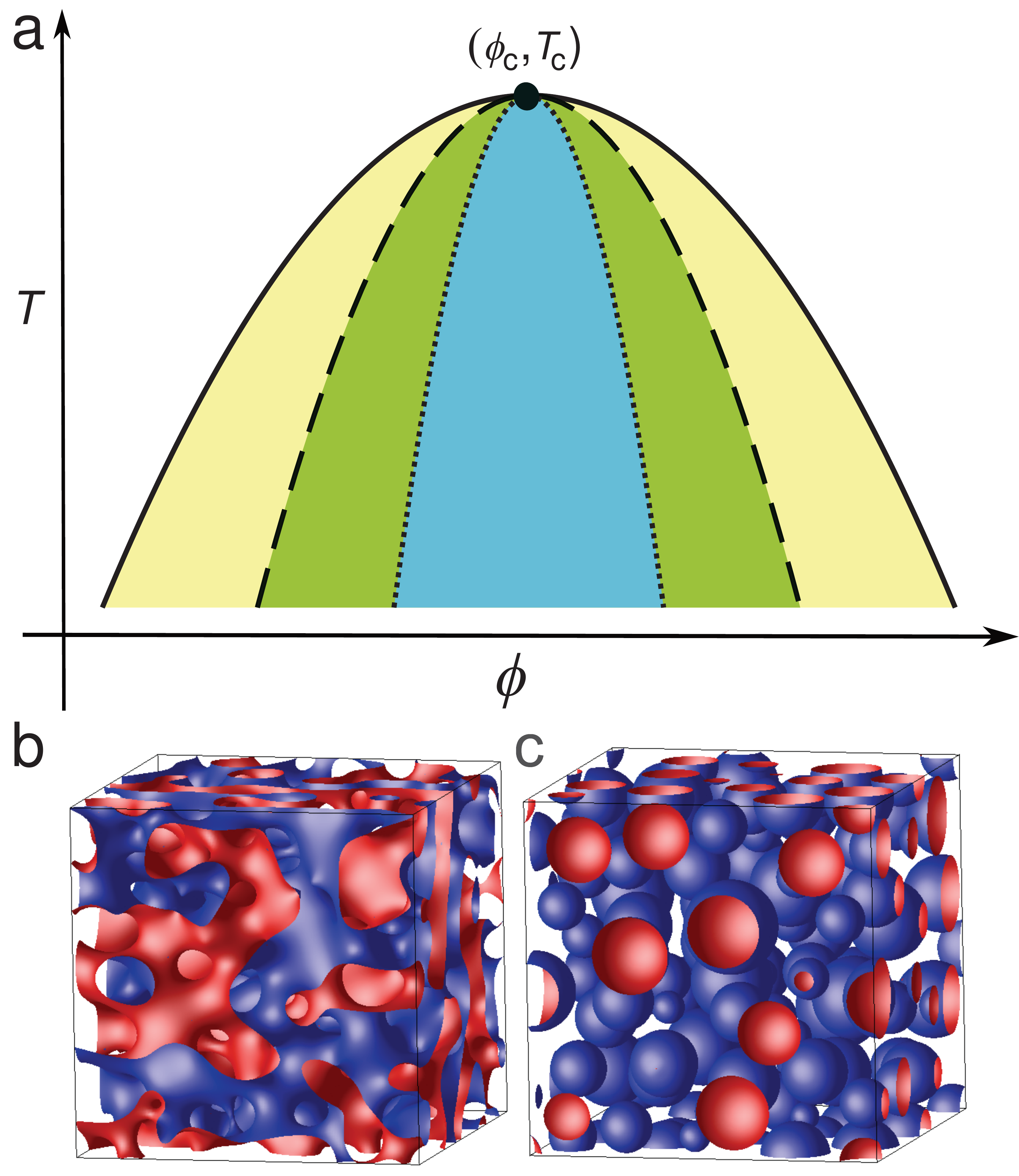}
\end{center}
\caption{{\bf A schematic phase diagram of a critical binary mixture with 
typical phase-separation structures.}
{\bf a,} Schematic phase diagram. In the (light yellow-green) metastable region ({\it i.e.}, the low $\Phi$ region ($\Phi<\sim 0.21$)), nucleation-growth-type phase separation takes place and there 
droplets coarsen by the evaporation-condensation (or, LSW) mechanism: diffusional transport from smaller to larger droplets. 
Near the symmetric composition (the light blue region, or, $\Phi \sim 1/2$), bicontinuous spinodal decomposition ({\bf b}) takes place and the domain coarsening is 
governed by the Siggia's hydrodynamic coarsening mechanism: hydrodynamic transport 
from thin to thick parts of bicontinuous tube structures. 
In the intermediate composition region (the green region), droplet spinodal decomposition ({\bf c}) takes place and its coarsening mechanism has been believed to be 
the Brownian coagulation mechanism: random Brownian motion of droplets and the resulting collisions. 
Contrary to this, we propose a new coarsening mechanism, in which droplet motion is induced by the 
composition Marangoni effect due to the diffusional composition coupling among droplets, and not by random Brownian motion.  
The border between bicontinuous and droplet spinodal decomposition is located around $\Phi=0.34$, according to our simulation. }

\label{fig:PD}
\end{figure}

Here we summarize the current understanding of liquid phase separation. We consider a binary liquid mixture, whose components have the same viscosity $\eta$. 
A schematic phase diagram is shown in Fig. \ref{fig:PD}a, together with typical phase-separation structures (Fig. \ref{fig:PD}b,c). 
The relevant domain coarsening mechanism depends solely on the volume fraction of the minority phase $\Phi$: For small $\Phi$ (case A), 
the evaporation-condensation (Lifshits-Slyozov-Wagner (LSW)) mechanism \cite{lifshitz1961,wagner1961theorie,voorhees1992ostwald} is responsible for droplet coarsening: steady diffusion flux from smaller droplets to nearby larger droplets leads to the growth of the latter at the expense of the former. In this case, the average domain size $\langle R \rangle$ grows as $\langle R(t) \rangle \cong [ (4/27) D_{0}\xi t ]^{1/3}$ \cite{onuki}, where $t$ is the time, $D_0$ is the diffusion constant of a component molecule (or atom), 
and $\xi$ is the correlation length of composition fluctuations (or the interface thickness). 
In this mechanism, translational diffusion of component molecules (or atoms) is the main transport process. 
Near a symmetric composition ($\Phi \sim 1/2$) (case B), bicontinuous spinodal decomposition takes place (Fig. \ref{fig:PD}b). 
There, the hydrodynamic coarsening (Siggia's) mechanism characteristic to a bicontinuous pattern leads to rapid coarsening \cite{siggia1979}: 
$\langle R(t) \rangle \cong 0.1 (\sigma/\eta)t$ \cite{onuki}, where  $\sigma$ is the interface tension between the two phases. 
In this mechanism, hydrodynamic flow is the main transport process.
For intermediate $\Phi$ (case C), droplet spinodal decomposition takes place (Fig. \ref{fig:PD}c). It has been believed that the Brownian-coagulation (Binder-Stauffer-Siggia (BSS)) mechanism is 
mainly responsible for the coarsening \cite{binder1974,siggia1979}: 
droplets grow by accidental collisions between droplets undergoing random Brownian motion, 
which leads to the coarsening law $\langle R(t) \rangle \cong [(6k_{B}T\Phi/(5\pi \eta)) t ]^{1/3}$ \cite{onuki,tanaka1996coarsening}, where $k_{\rm B}T$ is the thermal energy. 
In this mechanism, thermal Brownian motion of droplets (rather than molecules) is the main transport process.
The growth exponents of these coarsening laws have been confirmed 
by both experiments \cite{chou1979phase,wong1981light,hashimoto1986late,kuwahara1993,Perrot1994,beysens1997kinetics,bates1989spinodal} and 
numerical simulations \cite{shinozaki1993spinodal,koga1993late,kendon2001inertial,roy2013dynamics}. 

The coarsening mechanisms for case A and B are rather well established, whereas 
that for case C is not so clear due to 
difficulties associated with many-body effects and hydrodynamic effects. 
The simplifications of case A and B come from the fact that 
for case A transport is dominated by diffusion and for case B by hydrodynamic flow. 
On the other hand, both may play an important role for case C in a complex manner.
Although there has been a wide belief  that the primary coarsening mechanism is the BSS mechanism, 
additional mechanisms have also been suggested. For example, Tanaka 
observed the process of droplet coarsening in a binary liquid mixtures confined 
between two glass plates with optical microscopy, and found \cite{tanaka1994new,tanaka1996coarsening}  that 
droplet collision enhances subsequent collisions via hydrodynamic flow (phenomenon 1) and droplets sometimes move directionally, contrary to 
the expectation from random Brownian motion (phenomenon 2). 
Phenomenon 1 is enhanced for a quasi-two-dimensional (quasi-2D) situation due to stronger hydrodynamic interactions \cite{wagner2001phase}, but even for 3D  
its importance has been pointed out when the droplet volume fraction is very high \cite{martula2000coalescence,roy2013dynamics}. 
On phenomenon 2, some theories were proposed to explain directional motion of droplets 
by a coupling between composition and velocity fields \cite{tanaka1997new, kumaran2000spontaneous}; however, 
they turned out to be wrong even on a qualitative level, as will be described later. 
The difficulty of this problem stems  from the non-local nature of diffusion 
and the many-body and long-range nature of its coupling with hydrodynamic degrees of freedom. 
Although directional motion of droplets was experimentally reported for such a quasi-two-dimensional situation \cite{tanaka1994new,tanaka1996coarsening}, 
there has so far been no firm experimental evidence supporting such spontaneous motion of droplets for a bulk 3D system.  
Because of this lack of clear experimental evidence together with the theoretical difficulty mentioned above, phenomenon (ii) has not attracted much attention.  
Thus, it is now widely believed that droplet coarsening in immiscible liquid mixtures can be explained 
by the BSS mechanism \cite{gunton1983,binder1987theory,bray1994theory,onuki,puri2009kinetics,cates2012complex}. 

In this Article, we will demonstrate by numerical simulations that, contrary to the BSS mechanism, droplet motion is not random, but 
hydrodynamically driven by the composition Marangoni force, which is induced by long-range diffusional composition correlation among droplets, 
indicating that a dynamical coupling between the composition and velocity field leads to a novel  mechanism of efficient droplet coarsening in droplet spinodal decomposition. 

\vspace{0.3cm}
\noindent
{\bf Results}

\vspace{0.3cm}
\noindent
{\bf Phase diagram and the setting of basic parameters}

Before presenting results of phase-separation simulation, we summarize phase separation behaviour of a binary fluid mixture, 
focusing on how it depends on the composition $\phi$ and the temperature $T$ (see the schematic phase diagram, Fig. \ref{fig:PD}a).  
Here we consider phase demixing induced by an instantaneous temperature quench at time $t=0$ from an initial  temperature in the one-phase region (white region)
to a final temperature $T$ in the two-phase region. 
First the phase separation behaviour can be grouped into two types: nucleation\&growth (NG)-type phase separation in the metastable region (the light yellow region in Fig. \ref{fig:PD}a)
and spinodal-decomposition (SD)-type one in the unstable region. Then, SD-type phase separation can further be classified into 
droplet and bicontinuous SD. 
Domain coarsening is mainly driven by the LSW mechanism for NG-type phase separation, which takes place for $\Phi \ll1$, and by Siggia's mechanism for bicontinuous SD, which takes place for $\Phi \sim 1/2$. 
For droplet SD, which takes place for the intermediate $\Phi$ (roughly, $\sim 0.21<\Phi< \sim 0.34$; the green region in Fig. \ref{fig:PD}a), 
it has been widely believed that the BSS mechanism is responsible for domain coarsening.  
We note that the boundary between NG- and SD-type phase separation is sharp only in the mean-field limit, 
and should be diffuse for ordinary binary liquid mixtures \cite{tanaka1990transition}.  

The composition region on which we focus here is the region of droplet SD. 
The thermodynamic state of a binary fluid mixture is characterized by the composition $\phi$ and the temperature difference between the critical temperature $T_{\rm c}$ and the 
temperature $T$. Then the correlation length of the composition fluctuation $\xi$, 
which also characterizes the interface thickness, is described as $\xi \cong a |(T-T_{\rm c})/T_{\rm c}|^{-\nu}$, where the critical exponent $\nu \sim 0.63$ for 3D Ising 
universality class, to which a binary mixture belongs \cite{hohenberg1976,onuki}. On the other hand, the kinetics is characterized by the fluid viscosity $\eta$. Then  
the cooperative diffusion constant is expressed as $D_\xi=k_{\rm B}T/(6 \pi \eta \xi)$ and the characteristic lifetime of composition fluctuations 
is given by $\tau_\xi=\xi^2/D_\xi=6 \pi \eta \xi^3/k_BT$. 
As shown in Methods, the phase separation behaviour of a binary fluid mixture can be described solely by four dimensionless parameters. 
One is the scaled composition $\tilde{\phi}=\phi/\phi_e$, where $\phi_{\rm e}$ is the final equilibrium composition, 
or equivalently the volume fraction of the minority phase $\Phi$. 
The others are the fluidity parameter $\mathcal{A}$, which is a measure of the relative importance of the hydrodynamic to the diffusional transport  \cite{tanaka1998spontaneous}, 
the thermal noise parameter $\mathcal{B}$, which is the strength of the noise term (see Methods),  and the Reynolds number $Re$, which 
characterizes the relative importance of the inertia term to the viscous term.  
$\mathcal{A}$ and $\mathcal{B}$ are defined respectively as $\mathcal{A} =6\pi |\gamma| \xi^3 \phi_e^2= \frac{18 \pi \sigma \xi^2}{2^{3/2}k_{\rm B}T}$, 
and $\mathcal{B} =\frac{2}{|\gamma| \xi^3 \phi_e^2}= \frac{2^{5/2}}{3}\frac{k_{\rm B}T}{\sigma \xi ^2}$ \cite{koga1993late,tanaka1998spontaneous}.  
We note that $\sigma$ in these relations is the interface tension in an equilibrium two-liquid coexistence state (see Methods for 
its expression in the Ginzburg-Landau model).  We set $Re=0$ since in ordinary droplet phase separation we can safely neglect 
the inertia effects \cite{siggia1979} (see Methods).

We note that near a critical point $T_{\rm c}$, it is known that the renormalized value of $\mathcal{A}$ 
is a universal constant, $\mathcal{A}=4 \sim 8$ \cite{koga1993late,tanaka1998spontaneous}, since 
the equilibrium interface tension is given by $\sigma =(0.2 \sim 0.4) k_{\rm B}T/\xi^2$  
according to the two-scale-factor universality \cite{onuki}. 
However, far from a critical point, this parameter can become very large for typical immiscible binary liquids ({\it e.g.}, for a water/hexane mixture, $\mathcal{A} \sim 100$), 
whereas it becomes very small for polymer mixtures since $\mathcal{A} \sim N^{-1}$ ($N$: the degree of polymerization of polymer) \cite{onuki,koga1993late}. 
On the other hand, the estimation of the value of $\mathcal{B}$ needs some care, since this represents the strength of thermal fluctuations that 
directly lead to the break-down of the mean-field picture due to renormalization effects: A larger value of $\mathcal{B}$ shifts $T_{\rm c}$ to a lower 
temperature. 
We estimate that $\mathcal{B}$ should be the order of 1 near a critical point, and it decreases when the temperature is much lower than $T_{\rm c}$.  
So, in our simulations, we chose $\mathcal{A}=5$ and $\mathcal{B}=1$ as typical parameters for a system near a critical point. 
We note that the choice of these values of $\mathcal{A}$ and $\mathcal{B}$ does not affect our basic conclusion on a qualitative level. 
As shown below, our new mechanism, where hydrodynamic transport plays a key role, becomes more important with an increase in the distance from a critical point, 
{\it i.e.,} for larger $\mathcal{A}$ and smaller $\mathcal{B}$.  
Note that $\mathcal{A}/\mathcal{B}$ is a measure of the ratio of the hydrodynamic transport to the noise-driven diffusional transport.  
This means that our new mechanism should become more important for a liquid mixture far from a critical point, which is often the case for realistic applications, than for a critical mixture.    

Finally, we note that our study concerns only the so-called late stage of phase separation after the formation of a sharp domain interface, where 
the domain size is large enough compared to the interface thickness: $R \gg \xi$. 
Below we study $\Phi=0.25$, which is in the composition region of droplet SD (see Fig. \ref{fig:PD}a). 
Hereafter we use the scaled domain size $\tilde{R}=R/\xi$ and the scaled time $\tilde{t}=t/\tau_\xi$ to describe the coarsening dynamics. 

\vspace{3mm}
\noindent
{\bf Results without thermal noise}

\begin{figure*}[t!]
\begin{center}
   \includegraphics[width =15cm]{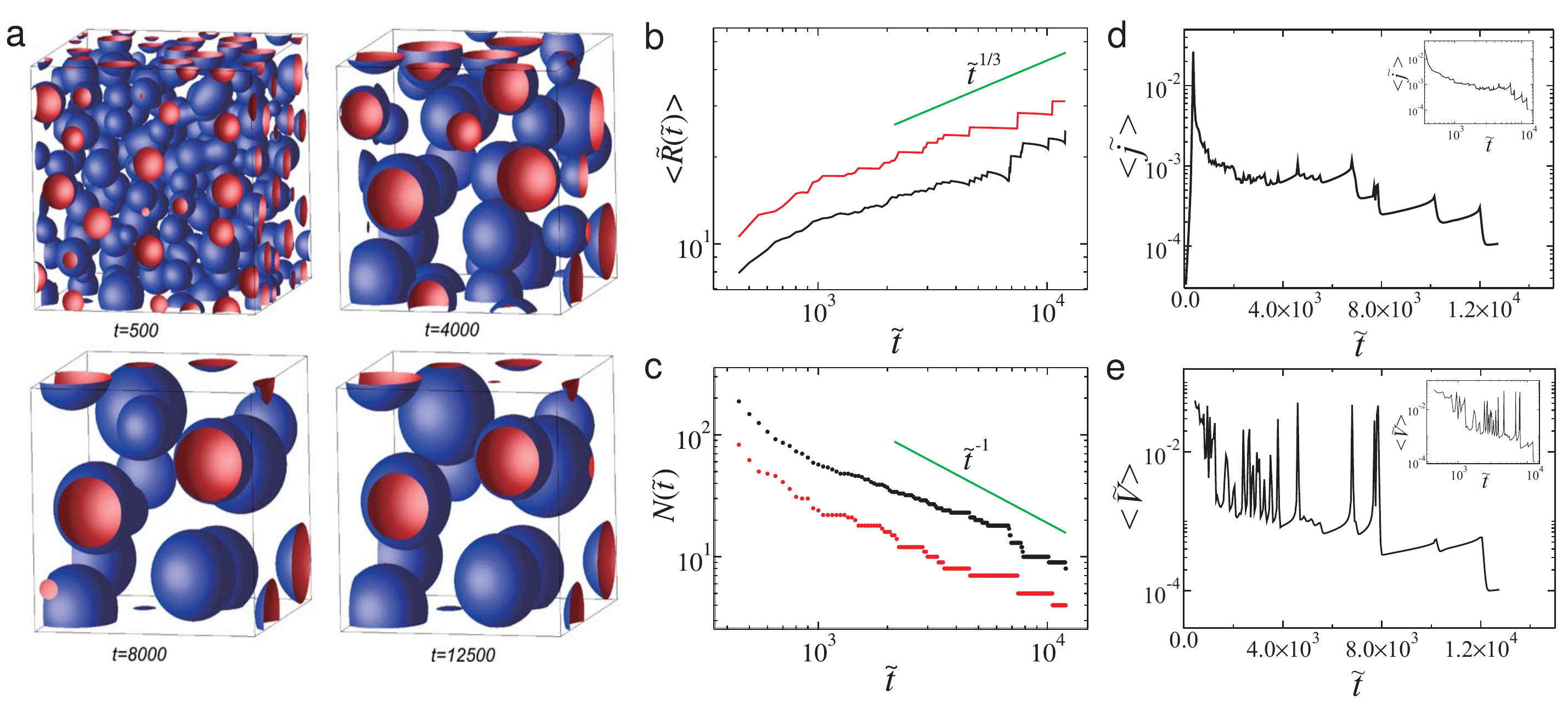}
\end{center}
\caption{{\bf Coarsening dynamics of droplet phase separation without thermal noise.} 
{\bf a,} Coarsening process of droplet spinodal decomposition at $\Phi=0.25$ for $\mathcal{A}=5$ and $\mathcal{B}=0$. 
Blue and red coloured surfaces show front and back sides of a droplet interface, respectively. 
Time evolution of the average droplet size $\langle \tilde{R}(\tilde{t}) \rangle$ and the total number of droplet $N(\tilde{t})$ are shown in {\bf b} and {\bf c} for $\mathcal{A}=5$ (black) and $\mathcal{A}=50$ (red), respectively. 
The green straight lines have the slope of 1/3 for {\bf a} and -1 for {\bf b}. 
 {\bf d,} Temporal change in the average diffusion flux $\langle \tilde{j} \rangle$ for $\mathcal{A}=5$. Note that the characteristic decay time of the diffusion flux after a collision is $\tau_\phi \sim 6 \pi \eta R^2a/k_{\rm B}T$. In our unit, this time scale is roughly proportional to $\tilde{R}^2$ (e.g., $\sim 400$ for $\tilde{R}=20$).   {\bf e,} Temporal change in the average velocity magnitude $\langle \tilde{V} \rangle$ for $\mathcal{A}=5$.   Note that the characteristic decay time of the velocity field   after a collision is $\tau_v \sim \eta R/\sigma$, which is the characteristic shape recovery time of a deformed droplet to a spherical shape.  In our unit, this time scale is roughly proportional to $\tilde{R}/\mathcal{A}$ (e.g., $\sim 4$ for $\tilde{R}=20$).  Each sharp spike in {\bf d} and {\bf e} reflects interdroplet collision.  The insets in {\bf d} and {\bf e} are the same plots but with a semi-log x axis. 
}
\label{fig:pattern_no_noise}
\end{figure*}

Firstly, we show a coarsening process during droplet phase separation in Fig. \ref{fig:pattern_no_noise}a, which is simulated by solving the dynamical equations of model H (see Methods) without thermal noise (see Methods). Because of the absence of noise ({\it i.e.}, $\mathcal{B}=0$), the droplet interface is smooth. 
The time evolution of the average size $\langle \tilde{R} \rangle$ and the total number of droplets $N$ for the fluidity  
parameter $\mathcal{A} = 5$ (black) and $\mathcal{A} = 50$ (red) are shown in Fig. \ref{fig:pattern_no_noise}b and c,
respectively. 
In addition to $\mathcal{A} = 5$, we use the large fluidity parameter $\mathcal{A} = 50$ to access a large separation between the interfacial thickness 
and the droplet size as well as to see hydrodynamics effects clearly. 
Because of the composition asymmetry, the symmetry of initial composition fluctuations around the average composition, 
which holds in the initial linear regime, is broken 
once the non-linear terms in the free energy start to come into play, and thus many small droplets are formed by fragmentation 
in the early stage. As stated above, in this study we are concerned only with the late stage of coarsening after the formation of a sharp domain interface.  
In the late stage, we find $\langle \tilde{R} \rangle \propto \tilde{t}^{1/3}$ and $N \propto \tilde{t}^{-1}$ for both fluidity parameters (Fig. \ref{fig:pattern_no_noise}b,c). 
Even when thermal noise is absent, we can see spontaneous motion of droplets and subsequent collisions: 
the droplet coarsening proceeds even without thermal noise.  
This result can never be explained by the Brownian coagulation mechanism, in which thermal noise is the only cause of droplet motion and coarsening.  
This is the most direct evidence for the presence of an unknown coarsening mechanism other than the BSS mechanism, 
which is responsible for the spontaneous motion of droplets and their collisions. 
Because of the absence of thermal noise, the droplet coarsening process is perfectly deterministic in this case, and thus the 
initial composition noise introduced at $t=0$ completely determines what takes place afterwards.

For $\mathcal{A}=5$, we can see both collisions of droplets and gradual disappearance of droplets due to evaporation, although the former is much more frequent than the latter. 
For $\mathcal{A} = 50$, on the other hand, the growth of droplets proceeds almost solely through direct collisions between droplets and evaporation of droplets is very rare. 
This difference can be explained by the fact that $\mathcal{A}$ is a measure of the relative importance of hydrodynamic to diffusional transport. 
Figures \ref{fig:pattern_no_noise}e and f show respectively the average diffusion flux and the average magnitude of the velocity of droplets for $\mathcal{A} = 50$ during a coarsening process. 
Each spike reflects interdroplet collision. 
We can see that, in the late stage, collisions take place after a complete decay of flow and thus are not affected by the preceding collision even for $\mathcal{A} = 50$, clearly indicating 
that the migration of droplets and the following collision are due to spontaneous motions of droplets and `not' induced directly by the preceding collision.

\vspace{3mm}
\noindent
{\bf Results with thermal noise}

Next we show a coarsening process during droplet phase separation with 
thermal noise in both composition and force fields. We note that the incorporation of both types of  
noise in the model H equations has been technically challenging \cite{camley2010dynamic,gross2010thermal,thampi2011lattice}. 
However, this has been properly done for 2D in the framework of model H by Camley et al. \cite{camley2010dynamic}. 
We have employed their method to simulate a 3D system (see Methods). 
The merit of this method is that we can drop the inertia term. 

\begin{figure*}[t!]
  \begin{center}
   \includegraphics[width =12cm]{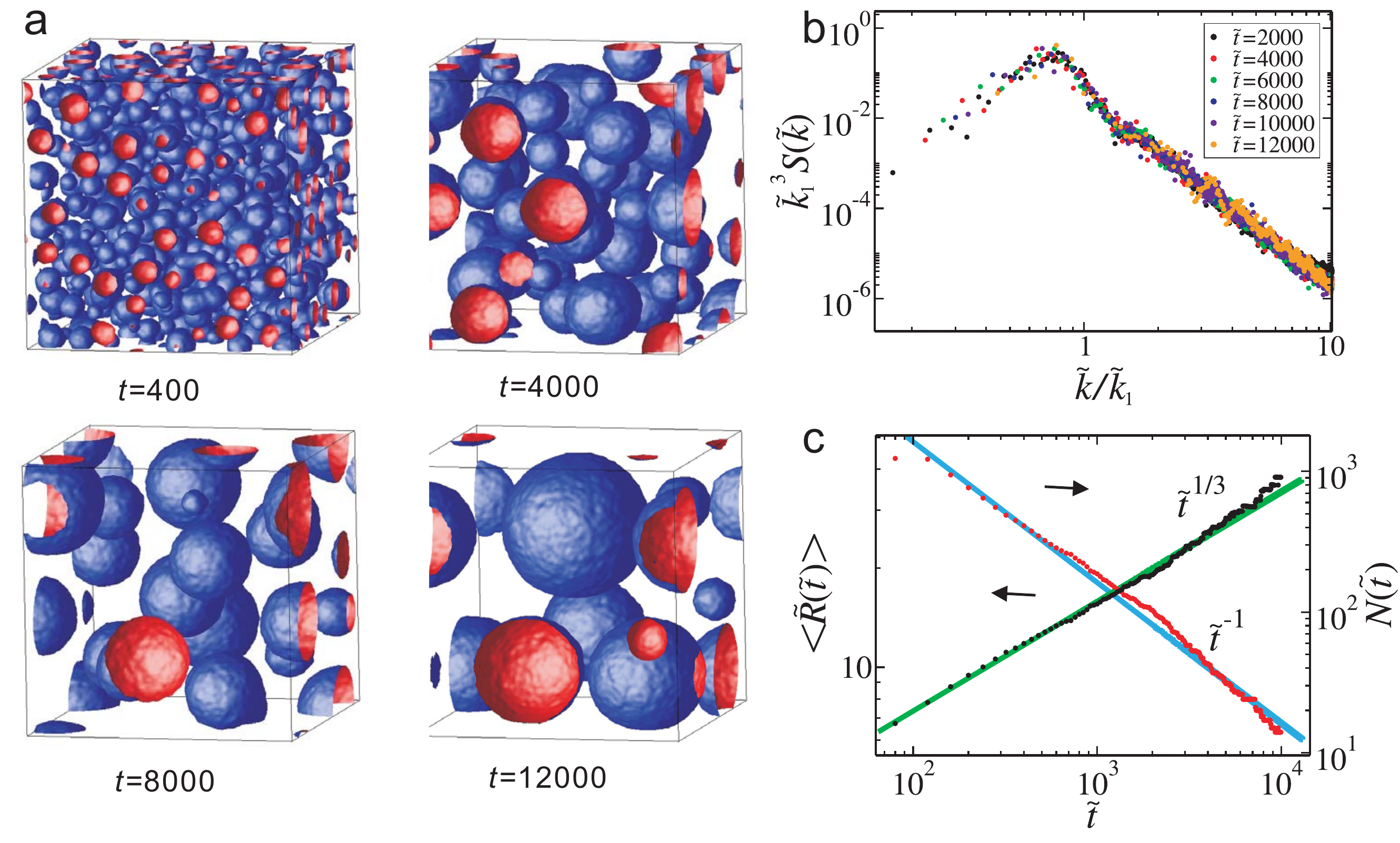}
  \end{center}
  \caption{{\bf Coarsening dynamics of droplet phase separation with thermal noise.} {\bf a,} Temporal change in 
  the droplet structure in the late stages of the phase-separation process at $\Phi=0.25$ for $\mathcal{A}=5$ and $\mathcal{B}=1$. 
  Blue and red coloured surfaces show front and back sides of a droplet interface, respectively. 
  {\bf b,}  Scaling plots of the normalised structure factor $S(k)$. The collapse onto the master curve indicates the self-similar nature of pattern evolution. 
  {\bf c,}  Time evolution of the average droplet size $\langle \tilde{R}(\tilde{t}) \rangle$ and the total number of droplets $N(\tilde{t})$   during phase separation. The blue and green lines have the  slope of -1 and 1/3, respectively. 
  }
  \label{fig:pattern_noise}
\end{figure*}

The process of phase demixing is shown in Fig. \ref{fig:pattern_noise}a (see also Supplementary Movie 1). 
The simulation is made for the fluidity parameter $\mathcal{A}=5$ and the thermal noise parameter $\mathcal{B}=1$ 
(see the above for the choices of these values). 
We can clearly see the temporal growth of droplets, which are induced by spontaneous droplet motion and the resulting collision and coalescence, as observed in Fig. \ref{fig:pattern_no_noise}a. 
Unlike in Fig. \ref{fig:pattern_no_noise}a, however, the interface of droplets are rough due to thermal fluctuation effects. 

As shown in Fig. \ref{fig:pattern_noise}b, the structure factor $S(\tilde{k})$ can be scaled by using the characteristic wavenumber $\tilde{k}_1$ (see Methods), indicating that 
droplet patterns grow self-similarly \cite{gunton1983,binder1987theory,bray1994theory,onuki,puri2009kinetics,cates2012complex}. 
The time evolutions of the average droplet size $\langle R \rangle$ and the number of droplets $N$ are shown in Fig. \ref{fig:pattern_no_noise}c. 
We find $\langle R \rangle \sim  t^{1/3}$ and $N \sim t^{-1}$. 
The exponents are consistent with the prediction of the BSS mechanism, where the coarsening rate is predicted as $\langle R(t) \rangle^3 \cong (6 \Phi k_{B}T/5\pi \eta)t$ \cite{onuki,tanaka1996coarsening}. 
In our scaled units, this relation becomes 
$\langle \tilde{R}(\tilde{t}) \rangle^3 \cong ((3 \Phi \mathcal{A} \mathcal{B})/5\pi) \tilde{t} \cong 0.24 \tilde{t}$ for $\mathcal{A}=5$ and $\mathcal{B}=1$. 
However, what we find is $\langle \tilde{R}(\tilde{t}) \rangle^3  \cong 1.6 \tilde{t}$. 
This indicates that the droplet size $\langle R \rangle$ grows 1.9 times faster than the prediction of the BSS mechanism although the exponent of the power law is the same. 
We note that our simulation result is much consistent with the experimental finding of Perrot et al. under microgravity \cite{Perrot1994} that the droplet size grows twice faster than the prediction of the BSS mechanism. This supports the validity of our simulations with the above choices of $\mathcal{A}$ and $\mathcal{B}$.

\vspace{3mm}
\noindent
{\bf Spontaneous directional motion of individual droplets}

\begin{figure}[t!]
  \begin{center}
   \includegraphics[width =8.5cm]{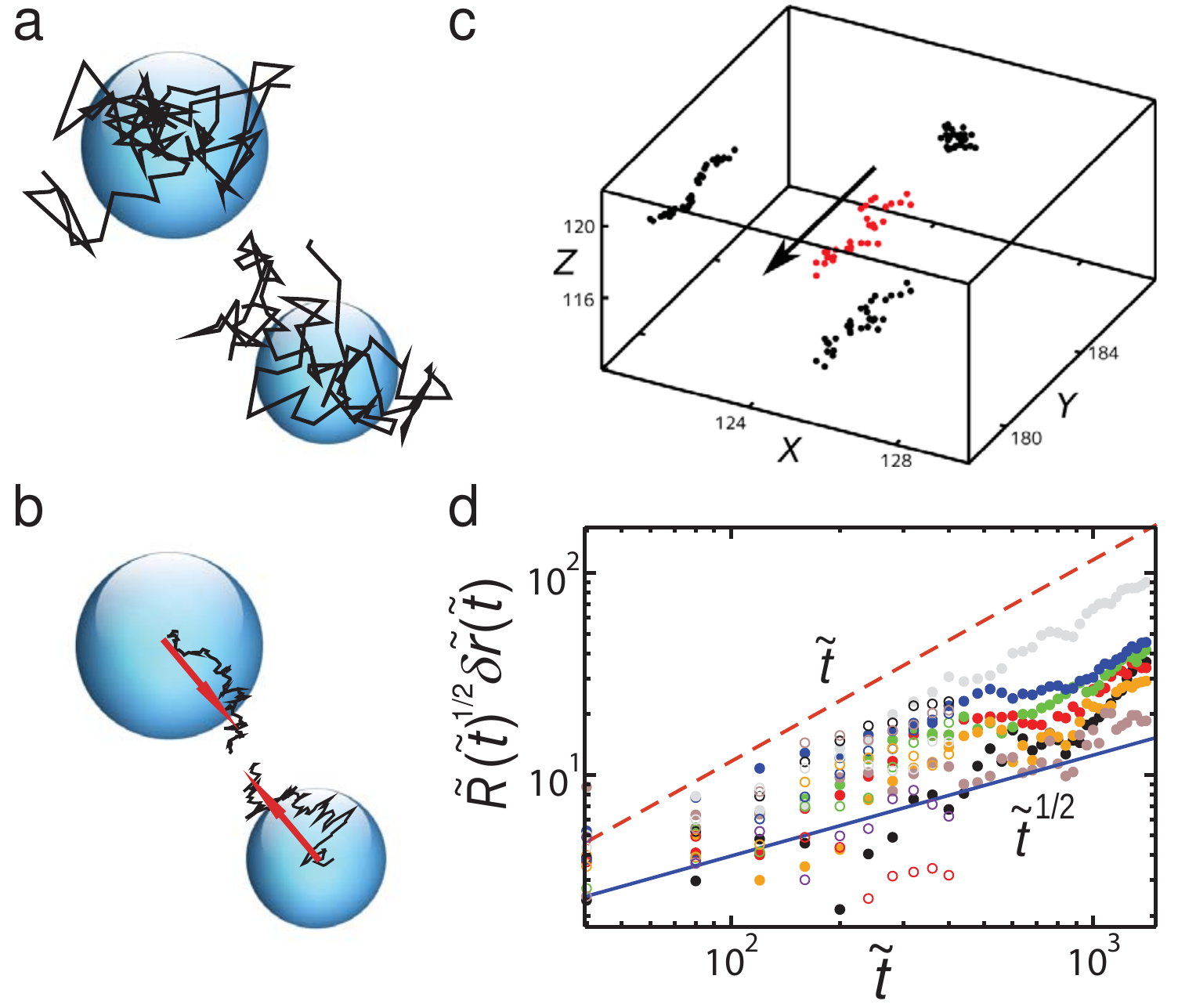}
  \end{center}
  \caption{{\bf Evidence for directional motion of droplets.} Schematic pictures for random and
   directional motion are shown in {\bf a} and {\bf b}, respectively. Case {\bf a} is for non-interacting 
   droplets and case {\bf b} for interacting droplets. 
  {\bf c:} An example of a trajectory of the centre of mass of a single droplet during $\tilde{t}=12120-13520$ in the process shown in Fig. 3. 
  The arrow indicates the direction of the motion.
  The result clearly shows directional translational motion in the phase-separating process, strongly suggesting 
  that the droplet motion is the type depicted in {\bf b} and not in {\bf a}.   
  {\bf d:} An example of displacements normalized by radius, 
  $\sqrt{\tilde{R}(\tilde{t})} \delta \tilde{r}(\tilde{t})$, of droplets during a time interval 
  (open symbols: $\tilde{t}=11320-11720$;  filled symbols: $\tilde{t}=12120-13520$) in the process shown in Fig. 3. We note that during this time interval no collisions take place. 
  Symbols with different colours are for different droplets in the simulation box. 
  The blue straight line is $\sqrt{\tilde{R}}\delta \tilde{r} \cong 0.4 \tilde{t}^{1/2}$ (see Eq. (1)) and the red dashed line has the slope of 1. 
  The displacements are always much larger than the prediction of Brownian motion (the blue straight line) for non-evaporating droplets. 
  Note that droplets having displacements smaller than the prediction are evaporating ones.
}
  \label{fig:trajectory}
\end{figure}

To seek the cause of the non-Brownian nature of droplet coarsening in simulations without noise (Fig. \ref{fig:pattern_no_noise}) and 
the faster coarsening than the BSS prediction in simulations with noise (Fig. \ref{fig:pattern_noise}), we analyse trajectories of droplets during the interval of two successive collisions 
for the case with thermal noise. 
First we focus on the nature of droplet motion, {\it i.e.}, whether it is completely random as assumed in the BSS mechanism (Fig. \ref{fig:trajectory}a) or directional 
(Fig. \ref{fig:trajectory}b). 
We carefully choose the observation period to avoid the effects of flow induced by collisions of droplets and to see 
only spontaneous motion of the droplets.
In Fig. \ref{fig:trajectory}c, a typical trajectory of a droplet during a certain time period are shown. We can clearly see that the droplet moves directionally 
(towards a neighbouring droplet). 
We confirm that this type of motion is common to most of droplets and not special (see, e.g., Fig. \ref{fig:trajectory}d). 
Below we will show that it is this directional motion that leads to the fast coarsening process. 
We note that this motion is not due to inertial effects \cite{roy2013dynamics} since our simulations are performed under the Stokes approximation ($Re=0$, or without the inertia term), 
which is valid for ordinary droplet phase separation. Thus, there must be some thermodynamic force acting on droplets that drives droplets to spontaneously move 
directionally.

For a droplet of radius $R$ which is undergoing random Brownian motion freely, its mean-square displacement, $\langle \Delta r^2 \rangle$, should be given by 
$\langle \Delta r (t) ^2 \rangle =\delta r(t)^2= (k_{B}T/(5\pi \eta R)) t$. In our scaled units, for $\mathcal{A}=5$ and $\mathcal{B}=1$, 
\begin{eqnarray}
\sqrt{\tilde{R}}\delta \tilde{r}=(\frac{\mathcal{A}\mathcal{B}}{10 \pi } \tilde{t})^{1/2} \cong 0.4 \tilde{t}^{1/2}. \label{eq:Brownian}
\end{eqnarray}
In order to compare the displacements of droplets with different radii, 
we plot $\sqrt {\tilde{R}}\delta \tilde{r}$ for all droplets in Fig. \ref{fig:trajectory}d. 
In this plot, the difference in the droplet size are scaled out; and, thus, the displacements of all droplets should agree to 
the prediction of Eq. (\ref{eq:Brownian})  and thus fall on the blue solid line in Fig. \ref{fig:trajectory}d, if droplets are doing free random Brownian motion. 
However, we find that most of droplets have much larger displacements far beyond the prediction of Eq. (\ref{eq:Brownian}) (or, the blue solid lines), 
strongly indicating the non-random directional nature of droplet motion.
During this time interval of observation, the change in the droplet radius is within a few \%, so this effect does not play any role.
Hydrodynamic interactions between droplets and finite-size effects should slow down the self-diffusion of droplets. 
Furthermore, the displacement should still increase in proportion to $\sqrt {t}$ as long as the motion is random. 
Nonetheless, Fig. \ref{fig:trajectory}d clearly show that the displacements of droplets are much larger than the prediction 
of Eq. (\ref{eq:Brownian}) and even increase almost linearly at later times. This cannot be explained by the scenario based on random Brownian motion. 

\begin{figure}[htbp]
  \begin{center}
   \includegraphics[width =8.5cm]{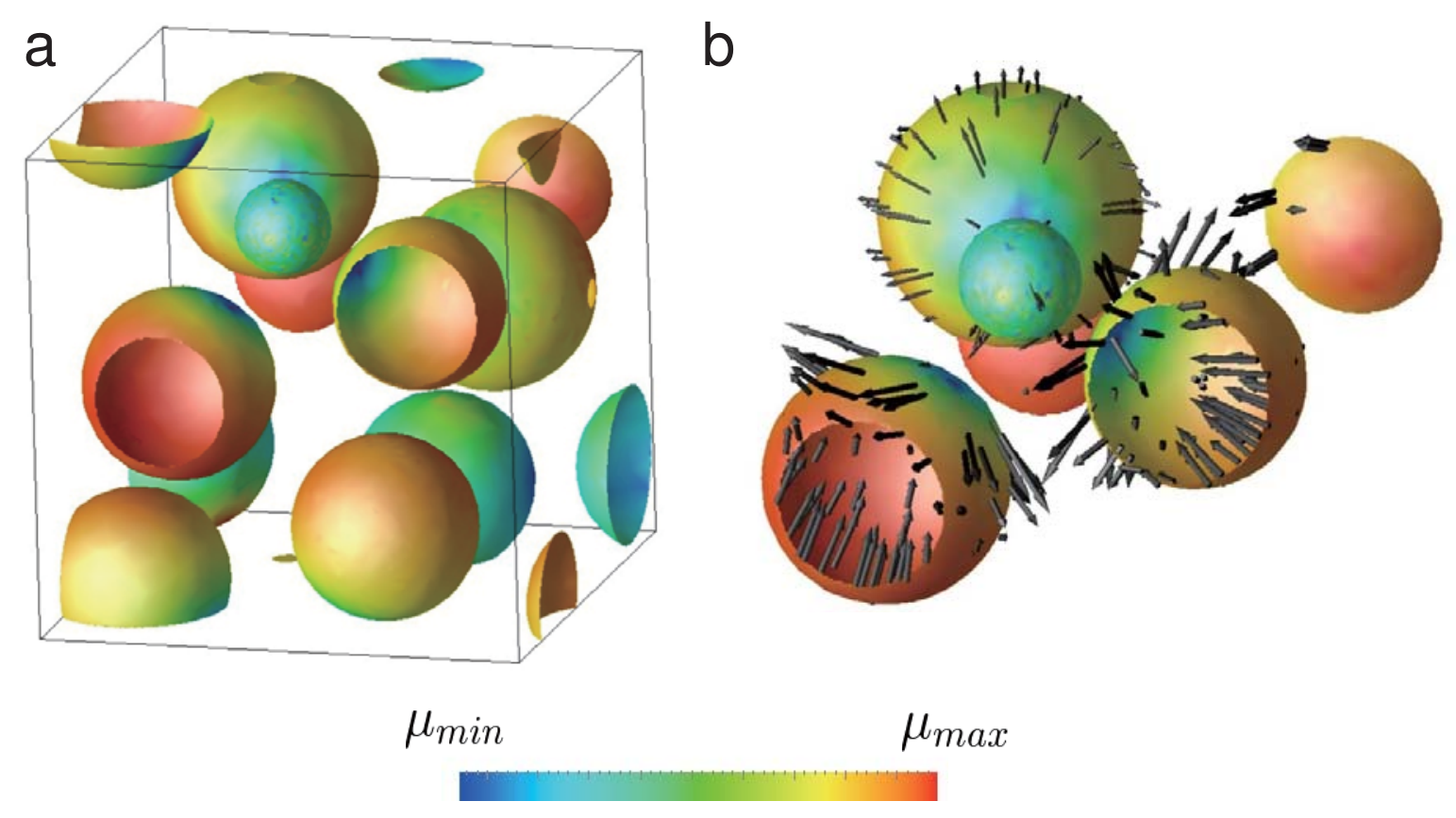}
  \end{center}
  \caption{{\bf Coupling between composition and velocity fields.} {\bf a,} A snapshot of
  droplets during phase separation without thermal noise. The colour reflects differences of the 
  chemical potential on the interface. Reddish and bluish colour means higher and lower chemical potential, 
  respectively. As shown in the colour bar, the colour is assigned for each droplet, which is bound by the minimum and maximum value of the chemical potential $\mu$. 
  This is because the amplitude of the change of $\mu$ inside each droplet interface is much smaller than the fluctuations of the absolute value of $\mu$ of each droplet. We note that the magnitude of difference relative to mean value is less than 3\% even for a droplet which has the largest intradroplet variation. 
  {\bf b,} Enlargement of the blue small droplet in {\bf a} and its surrounding droplets. The meaning of colour is the same as in {\bf a}.  
  The velocity field at the droplet interfaces is also shown by the arrows. 
  A longer arrow means high velocity. 
  }
  \label{fig:coupling}
\end{figure}

Without noise, we should be able to see clearly the relation between the thermodynamic force and the resulting transport, diffusional and hydrodynamic, 
without suffering from noise that obscures the details.  
Thus, we study the chemical potential on droplet interfaces for the case without thermal noise, and the result at a certain time is shown in Fig. \ref{fig:coupling}a. 
We can clearly see that the chemical potential on the interface of a droplet, or the interfacial composition profile, is not spherical symmetric but 
rather anisotropic, reflecting the spatial configuration of the neighbouring droplets.  As will be shown below, this causes the composition Marangoni force, which induces 
spontaneous directional motion of droplets. 
Nevertheless, the droplets are nearly spherical because of the action of the Laplace pressure $\Delta p=2\sigma/R$ , which induces a quick hydrodynamic relaxation to the stable spherical geometry  
with a characteristic relaxation time of $\tau_v \sim \eta R/\sigma$. 
A small variation of the interface tension, $\Delta \sigma$, in a droplet does not cause shape deformation of the droplet, but causes a drastic change 
in the type of droplet motion from stochastic to deterministic nature. 
We note that the former is a relative effect, whereas the latter is an absolute effect.  

Figure \ref{fig:coupling}b is an enlargement of the blue small droplet and its surrounding droplets shown in Fig. \ref{fig:coupling}a. 
The arrows show the flow field at their interfaces. 
The flow field is caused by an imbalance between the Laplace pressure and the thermodynamic force under the incompressibility condition: 
the break down of the spherical symmetry of the interfacial chemical potential in a droplet and the resulting tangential force acting along the interface allow 
the directional flow field to be induced even under the condition of $\mbox{\boldmath$\nabla$} \cdot \mathbf{v}=0$, 
unlike the case of an isolated spherical droplet.  
We can see that the smaller bluish droplet in a higher free energy state is slowly evaporating and the neighbouring droplets in lower energy states 
are growing. Thus the interface of a neighbouring droplet that is closer to the evaporating small droplet has smaller chemical potential (more bluish) 
than the opposite side. 
This anisotropy of the chemical potential on the interface of each droplet is the origin of the deterministic directional motion of the droplet. 
The variation of the chemical potential on a droplet interface is less than a few \%, but the resulting composition Marangoni force 
is strong enough to overwhelm the thermal force acting on a droplet.

The emerging physical picture is as follows. 
An small (evaporating) droplet causes deterministic flow fields on its neighbouring larger droplets, which lead to motion of the neighbouring droplets towards it and eventually 
induce an interdroplet collision, if the droplet does not evaporate before collision. 
In other words, a larger droplet tends to eat a neighbouring smaller droplet by direct collision. This hydrodynamic eating speed is much faster 
than the diffusional one in the evaporation-condensation (LSW) mechanism, as long as hydrodynamic transport is faster than diffusional transport.  
More precisely, however, the composition correlation between neighbouring droplets induces both the hydrodynamic and diffusional transport. 
Thus, the crossover between the present mechanism and the evaporation-condensation mechanism should be controlled by the competition between hydrodynamic 
and diffusional transport. 
The former is always dominant as long as $\Phi$ is not so low. 
For lower $\Phi$, however, the direct compositional correlation between droplets, which is the origin of the 
thermodynamic force inducing hydrodynamic droplet motion, becomes weaker  
due to the less overlap of the composition fields around droplets. For very low $\Phi$, thus, droplets interact with each other 
via the mean-field matrix, as assumed in the original mean-field theory of the LSW mechanism \cite{lifshitz1961,wagner1961theorie}. 
In this regime, the LSW mechanism becomes dominant. 
We speculate that the crossover takes place around $\Phi \sim 0.21$ separating NG and SD  (see Fig. \ref{fig:PD}) since 
by definition droplets are formed rather independently for NG whereas in a correlated manner for SD. But this crossover composition may also depend on the value of $\mathcal{A}$. 
Although this is a very interesting problem, we leave it for future investigation.

\vspace{3mm}
\noindent
{\bf Analysis of a pair of droplets}

\begin{figure*}[htbp]
  \begin{center}
   \includegraphics[width =16cm]{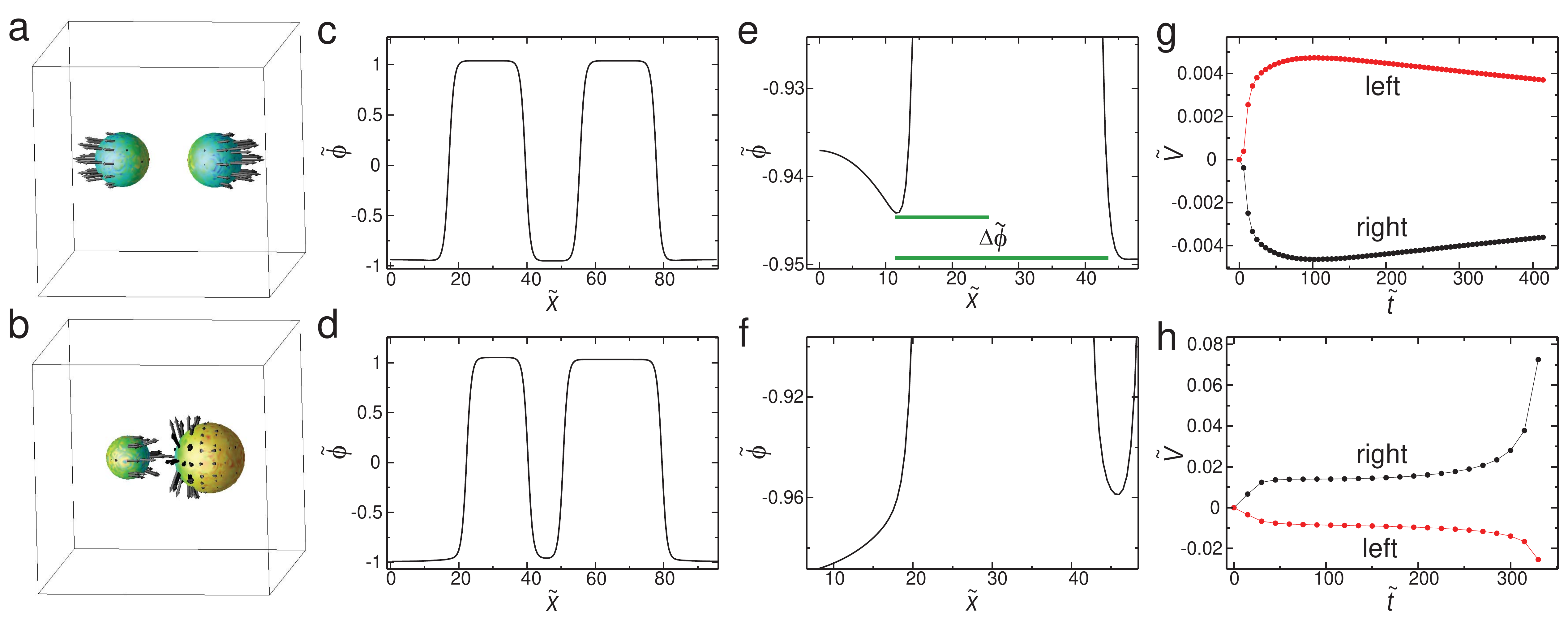}
  \end{center}
  \caption{{\bf Motion of a pair of growing droplets.} {\bf a,} Motion of two growing droplets of the same size. 
  The centre-of-mass distance between the two droplets is 35 and the initial radii of the two droplets are both 10. The 
  velocity field and the chemical potential at the droplet interfaces are also shown, as in Fig. \ref{fig:coupling}. 
  The interfacial tension of the left droplet is 0.9286 for its left side and 0.9296 for the right side around $\tilde{t}=100$. The theoretically estimated velocity 
  from these values is -0.0050, which is consistent with the measured velocity.     
  The colour bar is the same as in Fig. \ref{fig:coupling}.
  We can see the higher interface tension at the side of a droplet near another one. 
  This leads to the motion of the droplets increasing their separation, 
  due to composition Marangoni effects. 
  {\bf b,} Motion of a pair of droplets with different size.  The centre-of-mass distance between the two droplets is 45, and the initial radii of the left and right   
   droplets are 10 and 15 respectively. The sides of the droplets nearest each other have lower interfacial tension and the two droplets approach each other. 
   The interface tension of the left and right side of the left droplet is 0.9303 and 0.9294 respectively and the velocity estimated from these values is +0.0045, whereas the interface tension  of the left and right side of the right droplet is 0.9289 and 0.9313 respectively and the velocity estimated from these is -0.012, around $\tilde{t}=100$. These velocities estimated around $\tilde{t}=100$ are consistent with the measured ones. 
  The simulation box size is 192$^3$ for both {\bf a} and {\bf b}. Because of its small size, the results suffer from finite-size effects. 
  {\bf c,d:} The composition profile of the droplets along the line connecting their centres of mass for {\bf a} and {\bf b}, respectively. 
  {\bf e,f:}  Enlargement of the composition profile of the left droplet shown in {\bf c} and {\bf d}, respectively.  
  {\bf g,h:} The temporal change of the centre-of-mass velocity of the two droplets for {\bf a} and {\bf b}, respectively. 
  The final steep change of the velocity in {\bf g} is due to a direct overlap of the interfaces of the two droplets upon collision.
  }
   \label{fig:pair}
\end{figure*}

To elucidate the exact physical mechanism, we also study the motion of pairs of droplets (see Methods). 
Typical results are shown in Fig. \ref{fig:pair}a and b, where the chemical potential and the flow field on the interfaces are shown as in Fig. \ref{fig:coupling}b. 
Figure  \ref{fig:pair}c and d show the composition profile of two droplets along the line connecting the centres of mass of the two droplets 
for case a and b.  Figure \ref{fig:pair}e and f are their enlargements, 
which clearly show the presence of the composition difference by $\Delta \tilde{\phi}$ between the two sides of a droplet. 
We can see that such an asymmetric interfacial profile of a droplet induces a chemical potential gradient 
in the droplet and leads to spontaneous directional droplet motion. Finally, the temporal change in the centre-of-mass velocities of the two droplets are shown in 
Fig. \ref{fig:pair}g and h, respectively, for case a and b.

For the case of Fig. \ref{fig:pair}a, we calculate the interfacial tension $\sigma $ for each interface by integrating 
$\mathbf  F_{\phi } = -\nabla ^2 \tilde{\phi} \nabla \tilde{\phi} $ \cite{tanaka1997new} across the interface along the radial direction. 
A part of this thermodynamic force is balanced by the Laplace pressure $\Delta p=2 \sigma/R$. 
For the left droplet whose radius is 11.5 and velocity is -0.0045, the interfacial tension is 0.9286 for its left side and 0.9296 for the right side.  
This small variation of the interfacial tension is consistent with a nearly spherical shape of droplets, as discussed above. 
It is the interfacial tension gradient due to the composition gradient on the interface that leads to the motion of a droplet. This is known as the composition Marangoni effect  
 \cite{young1959motion,vladimirova1999diffusiophoresis}. 
We note that it is well known that this effect induces spontaneous droplet motion for droplets with weak surfactants \cite{golovin1995spontaneous}  
and those subjected to composition or temperature gradients \cite{jayalakshmi1992phase,enomoto2002coarsening}. 
We stress that unlike these well-known cases, for the present case the interfacial tension gradient in a droplet 
is produced by a non-trivial coupling between the composition fields around droplets.  
The analytical expression for the droplet velocity under a uniform composition gradient is give as $\mathbf{V} = -(R/(5\eta)) (d \sigma/dx)$ \cite{young1959motion}. 
By inserting to this relation the above values of $\sigma$ estimated from the simulation, we estimate the droplet velocity as $|\mathbf{V}| \sim 0.0050$, which is much consistent with the directly measured one ($|\mathbf{V}| \sim 0.0045$) (see Fig. \ref{fig:pair}g). 
A slight difference may come from the difference in the geometry, {\it i.e.}, a single particle under a uniform composition gradient (theory) vs. 
a droplet pair under diffusional coupling (our simulation), and the finite size of the simulation box. 
To avoid the finite-size effects, we need to perform larger size simulations, but unfortunately a large numerical cost of hydrodynamic simulations 
do not allow us to do so.  
The velocities of the droplets in Fig. 7b are also estimated from the same analysis, and they also 
agree well with the measured ones (see Fig. 7b and its caption). These results strongly support our mechanism.

\begin{figure}[htbp]
  \begin{center}
   \includegraphics[width =6.5cm]{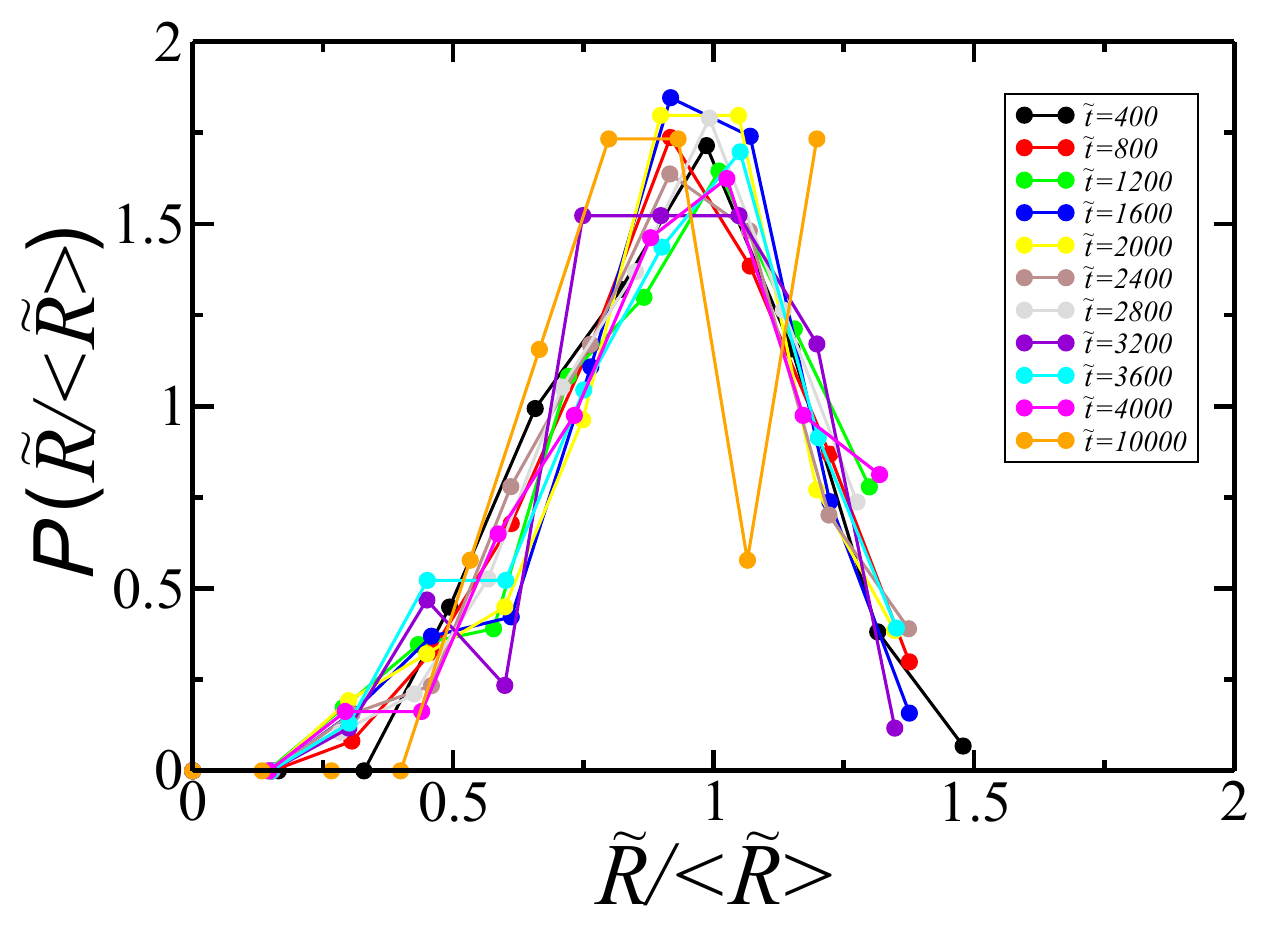}
  \end{center}
  \caption{{\bf The temporal change in the normalized droplet size distribution function.} 
  We can see that the normalized droplet size distribution functions at various phase-separation times can be scaled onto a universal curve at least approximately, 
  consistent with the self-similar nature of the domain growth (see Fig. 3b).  
  }
  \label{fig:distribution}
\end{figure}

In the particular configuration in Fig.  \ref{fig:pair}a, due to the perfect symmetry around the centre of mass of the two droplets and the absence of noise, coarsening does not proceed. 
We note that the same situation should be realized for the LSW mechanism. 
This is not the case for the asymmetric situation in Fig.  \ref{fig:pair}b, where two particles approach each other and eventually collide. 
For real droplet phase separation, the randomness of droplet size and many-body interactions always lead to such asymmetric situations and their continuous presence 
is responsible for continuous coarsening. 
We emphasize that the randomness is a direct consequence of thermal fluctuations and thus intrinsic to any phase separation phenomena.
Combining this result on pairs of droplets and the results shown in Fig. \ref{fig:coupling}a and b, we can conclude that the composition Marangoni effect  
stemming from the randomness of droplet size is responsible for spontaneous directional motion of droplets and the resulting coalescence. 
The importance of randomness indicates that there should be a non-trivial feedback between the coarsening and the droplet size distribution. 
We find that the normalized droplet size distributions at various phase-separation times can be superimposed on the universal curve (see Fig. \ref{fig:distribution}), 
consistent with the self-similar nature of pattern evolution. 
We speculate that more frequent collisions between particles with large size difference may be responsible for the self-similarity, but  
this problem should be studied carefully in the future. Our mechanism cannot lead to domain coarsening for a mono-disperse emulsion. 
However, the Brownian coagulation mechanism should work even in such a situation; and once a collision happens and produces the droplet size difference,  
our mechanism should become operative. 

\vspace{3mm}
\noindent
{\bf Roles of other non-trivial mechanisms}

Finally we briefly consider other possibilities. 
In our previous papers \cite{tanaka1994new,tanaka1996coarsening}, we suggested two other non-trivial droplet collision mechanisms: 
multiple-collision and collision-induced collision mechanisms, which was confirmed experimentally and numerically \cite{martula2000coalescence,wagner2001phase,nikolayev1996new}.  
Here we show that neither of them can explain the above-described phenomena. 
The former mechanism is a rather obvious one. When two droplets collide with each other, the shape of the droplet interface is strongly deformed and 
relaxes to a spherical shape, accompanied by the interfacial-tension driven flow. This flow induces hydrodynamic motion of the surrounding droplets, which 
leads to subsequent collisions. 
The latter mechanism is, on the other hand, a result of a coupling between droplet collision and composition diffusion. After droplet collision, strong diffusion flux towards 
the resulting merged droplet is created due to what we call the interface quench effect \cite{tanaka1998spontaneous} (see Fig. \ref{fig:pattern_no_noise}d), 
which is the process of local equilibration of the droplet with the matrix. 
This strong diffusion affects the translational motion of other neighbouring droplets due to the dynamical coupling of the diffusion and flow fields, which leads to subsequent collisions. 
Both mechanisms are expected to enhance successive collisions, but are effective only immediately 
after previous collisions. 
To reveal whether the coarsening behaviour can be described by these mechanisms or not, 
we calculate the average diffusion flux and the average magnitude of the velocity during a coarsening process without noise respectively in Fig. \ref{fig:pattern_no_noise}d and e. 
Occasional interdroplet collisions accompany sharp spikes and subsequent relaxation for both quantities. 
The characteristic decay times of diffusion flux and flow are given by $\tau_\phi=R^2/D_0=6 \pi \eta R^2 a/(k_{\rm B}T)$ and 
$\tau_v=\eta R/\sigma$, respectively.
As can be seen in Fig. \ref{fig:pattern_no_noise}d and e, in the late stage a collision takes place after the diffusional and hydrodynamic fluxes induced by 
the preceding collision have completely decayed, suggesting that it is not affected directly by 
the preceding collision: there are no chain collisions.  
This strongly indicates that the coarsening behaviour we observe in the late stage of phase separation can be explained by neither of the above-mentioned mechanisms.

\vspace{3mm}
\noindent
{\bf Discussion}

The dynamical scaling argument on domain coarsening in the late stage of phase separation has been so successful \cite{gunton1983,binder1987theory,bray1994theory,onuki,puri2009kinetics,cates2012complex}. In this argument, it has been assumed that 
the compositions of the two phases already reach the final equilibrium one $\phi_{\rm e}$ in the late stage. 
For droplet phase separation, thus, it is implicitly assumed that the interfacial tension $\sigma$ is uniform over a droplet's interface. 
In liquid mixtures, which are incompressible in ordinary conditions, the hydrodynamic flow can be induced only by the interfacial force,  
which acts normal to the interface in the above assumption. 
For a spherical droplet, this leads to the conclusion that the spherical symmetric force fields which act along the interface normal unit vector $\mathbf{n}$ of 
a droplet cannot cause any flow under the constraint of 
$\mathbf{\nabla} \cdot \mathbf{v}=0$ and accordingly the interfacial force should be balanced by the Laplace pressure $\Delta p=2\sigma/R$ 
($\sigma $ being the interfacial tension) \cite{tanaka2000viscoelastic}. 
This is the basis for the long belief that droplet motion is not induced hydrodynamically but solely by its thermal diffusion. 

In the above, we have shown that this physical picture is not correct and the interfacial tension is actually inhomogeneous over each droplet's surface due to diffusional 
composition correlation among droplets, which allows interfacial-tension-driven flow even for a spherical droplet. 
The key is the fact that droplets exchange the component molecules (or atoms) with neighbouring droplets via the surrounding matrix phase by diffusion. 
The time necessary for a component molecule (or atom)  (size $a$) to diffuse over the same distance is $\tau _{s}= L^2/D_0$. 
On the other hand, the average time necessary for a droplet of radius $R$ to diffuse over an average inter-droplet distance 
$L$ is estimated as 
$\tau _{D}= L^2/D_{R}$, where $D_{R}$ is the diffusion constant of a droplet.  
Since $D_0$ and $D_{R}$ are inversely proportional to the size of a component molecule (or atom), $a$, and the domain size, $R$, respectively, 
the ratio $\tau _{D}/ \tau _{s}$ becomes very large in the late stage of droplet SD, where $R \gg a$. 
Thus the composition correlation between neighbouring droplets develops via molecular (or atomic) diffusion long before 
an accidental collision between them by thermal Brownian motion takes place. 
It is this composition correlation among neighbouring droplets that is responsible for internal inhomogeneity of the interfacial tension over 
a droplet's interface and the resulting Marangoni force.  

Here we note \cite{siggia1979, tanaka1996coarsening} that van der Waals attraction between droplets is too weak to affect the motion of droplets 
 in the late stage. Furthermore, directional motion of droplets observed in our previous experiments in a quasi-2D situation, where neither gravitational force 
 nor electrostatic interactions are relevant, can be explained by the present mechanism. 

Here it is worth mentioning that the mechanism we find here is essentially different from 
mechanisms previously proposed by one of the authors (H.T.) \cite{tanaka1997new} and Kumaran \cite{kumaran2000spontaneous}. 
In the above, we reveal that the force acting on a droplet comes from the non-uniformity in the amount of the composition jump across the interface in the droplet  
and thus is localized on the interface, whose characteristic width is given by $\xi$. 
This non-uniformity of the interfacial tension due to the coupling of the composition fields around droplets 
causes the composition Marangoni flow, which leads to the spontaneous motion of droplets. 
In both of the previous theories, however, the non-uniformity of the Gibbs-Thomson condition on the droplet interface was completely ignored 
and only the composition gradient in the matrix phase was considered. 
The characteristic lengthscale of this inhomogeneous composition is the order of $R$. 
Thus, the strength of the composition Marangoni force in the new mechanism is much stronger than the composition-gradient-induced force in the previous ones 
since $1/\xi \gg 1/R$. Furthermore, there is another important, even qualitative, difference between them; 
for example, both of the previous theories  \cite{tanaka1997new,kumaran2000spontaneous} (wrongly) predict 
that for a pair of large (growing) droplets or for  a pair of small (shrinking) droplets, the two droplets should come towards each other, 
whereas for a pair of large (growing) and small (shrinking) droplets the two droplets should move away from each other. 
We stress that this is opposite to what is shown in Fig. \ref{fig:pair}a and b.  

The novel mechanism reported here can be explained on an intuitive level as follows. 
There is a difference in the local free energy between droplets, reflecting the difference in the size between them. 
Larger droplets are more stable and in a lower free energy state simply because the interface energy cost is smaller. 
This leads to the diffusion flux from small droplets to nearby larger droplets due to the chemical potential difference. 
This correlation of the composition field between droplets evolves much faster than any significant diffusional centre-of-mass 
motion of the droplets. 
This is exactly the same process as that of the LSW mechanism: if there are neither hydrodynamic degrees of freedom nor translational droplet diffusion, this leads to the coarsening of the LSW type.     
However, the presence of hydrodynamic degrees of freedom in a system entirely changes the kinetic route for lowering the free energy. 
An intradroplet gradient of the interfacial tension generates hydrodynamic motion of droplets towards the side with  
lower interfacial tension. Thus, larger droplets tend to eat nearby small droplets, as in the case of the LSW mechanism. However, the crucial difference comes from the fact that 
the transport process is dominantly controlled by hydrodynamic translational motion of droplets due to the composition Marangoni effect in our mechanism, whereas by translational diffusion of 
molecules (or atoms) in the LSW mechanism.

In a more general perspective, we may say that domain coarsening is 
under the influence of the free energy at any moment. This has a significant implication for the unified physical picture of domain coarsening: all coarsening mechanisms, including the evaporation-condensation mechanism (for $\Phi \ll 1$), the hydrodynamic coarsening mechanism ( for $\Phi \sim 1/2$), and the mechanism found here (for intermediate $\Phi$), 
have a common principle of material transport: Transport can be either diffusion, flow, or by their coupling, but always takes place from smaller to larger domains: Domain coarsening obeys a rule that ``big always wins over small''. 
We note that there is no such a rule for the BSS mechanism, in which the process is purely random and stochastic.

Finally, we stress that our new mechanism may play an important role in phase separation of many binary mixtures in practical use, 
which takes place in a state far from a critical point. We note that our mechanism should be more operative for a mixture of lower viscosity, as mentioned earlier. 
We also stress that our mechanism can operate even near zero temperature, as shown in Fig. \ref{fig:pattern_no_noise}, unlike the Brownian-coagulation mechanism; 
and thus it should be relevant to spinodal decomposition in quantum systems, which often takes place at extremely low temperatures \cite{Hemixture,BEC,BEC2}. We note that 
in such a condition, quantum diffusion associated with tunnelling mass transfer can be operative but there is little thermal motion of droplets. 
Our study shows that the coupling between the composition and velocity fields alone can lead to domain coarsening even with little thermal noise. The relative importance of the velocity field to the composition field monotonically increases with an 
increase in the volume symmetry between the two phases; and thus the primary mechanism of coarsening changes from the LSW, to our mechanism, and to the Siggia's mechanism 
with an increase of $\Phi$ from 0 to 1/2. 
We hope that our mechanism will provide a novel perspective for spinodal decomposition of classical and quantum fluid systems in various fields.

\vspace{2cm}
\noindent
{\bf Methods}

\noindent
{\bf Basic dynamic equations of model H.}
For our simulations, we employ the following kinetic equations known as a fluid model, or model H \cite{gunton1983, hohenberg1976, onuki}: 
\begin{equation}
\frac{\partial \phi }{\partial t} =-\mathbf v \cdot \nabla \phi + L\nabla ^2\frac{\delta(\beta \mathcal{H})}{\delta \phi }  + \theta, \label{phiflux}
\end{equation}
\begin{equation}
 \rho \frac{\partial \mathbf v}{\partial t} + \rho (\mathbf v \cdot \nabla )\mathbf v = \mathbf F_{\phi }-\nabla p + \eta \nabla ^2 \mathbf v  + \mbox{\boldmath$\zeta$}, \label{ns}
\end{equation}
where $\phi $ is the composition, $\mathbf{v}$ is the fluid velocity, $p$ is a part of pressure, $\rho $ is the density, $\eta $ is the viscosity, $L$ is the kinetic coefficient, 
and $\beta=1/k_{\rm B}T$ ($k_{\rm B}$: the Boltzmann constant). 
The pressure $p$ is determined to satisfy the incompressiblity condition $\mathbf \nabla \cdot \mathbf{v}=0$. 
Since our primary concern is the dynamical coupling between the composition and flow field, we do not care  about the detailed properties of the mixture and thus 
adopt the following standard Ginzburg-Landau free energy functional: 
\begin{equation}
\beta \mathcal{H} = \int d\mathbf{r} \ [\frac{\gamma}{2}\phi ^2+\frac{u}{4}\phi ^4+\frac{K}{2}|\nabla \phi |^2], 
\end{equation}
where $\gamma=\gamma_0 (T-T_{\rm c})$ and $\gamma_0$, $u$ and $K$ are positive constants. 
In Eq. (\ref{ns}), $\mathbf F_{\phi }$ is the thermodynamic force density acting on the fluid due to  spatial inhomogeneity of the composition field 
and $\mathbf F_{\phi } = -\phi \nabla \mu = -\nabla \pi + k_{\rm B}TK\phi \nabla ^2\nabla \phi$ ($\pi$: osmotic compressibility), 
where $\mu = \delta \mathcal{H}/\delta \phi$ is the chemical potential. Here $\theta$ and $\mbox{\boldmath$\zeta$}$ are the thermal noise terms in composition and force, which satisfy the following fluctuation-dissipation relation, respectively \cite{gunton1983,onuki}.
\begin{equation}
   \langle \theta (\mathbf  r,t)\theta (\mathbf r^{'}, t^{'}) \rangle = -2L\nabla ^2\delta (\mathbf  r-\mathbf  r^{'})\delta (t-t^{'}), \label{noise_phi}
\end{equation}
and
\begin{equation}
   \langle \zeta _{i}(\mathbf  r,t)\zeta _{j}(\mathbf r^{'}, t^{'}) \rangle = -2k_{\rm B}T\eta \delta _{ij}\nabla ^2 \delta (\mathbf  r-\mathbf  r^{'})\delta (t-t^{'}). \label{noise_stress}
\end{equation}

Here we rewrite Eqs. (\ref{phiflux}) and (\ref{ns}) by scaling the space and time unit by the correlation length of critical composition fluctuations 
$\xi = (K/|\gamma|)^{1/2}$ and the characteristic lifetime of fluctuations $\tau _{\xi } = \xi ^2/(|\gamma| L)$. Then we define new variables, $\tilde{\mathbf r} = \mathbf{r} /\xi $ ($\mathbf {r}$ being the position vector), 
$\tilde{t}= t/\tau _{\xi}$  and $\tilde{\mathbf v} = (\xi/\tau) \mathbf{v}$. The composition is normalized as $\tilde{\phi} = \phi /\phi _{\rm e}$, where $\phi _{\rm e} = (|\gamma|/u)^{1/2}$ is the equilibrium composition of the coexisting phase. The scaled equation which corresponds to Eq. (\ref{phiflux}) is then 
\begin{equation}
\frac{\partial \tilde{\phi}}{\partial \tilde{t}} =-\tilde{\mathbf{v}} \cdot \nabla \tilde{\phi} 
+ \nabla ^2[-\tilde{\phi} + \tilde{\phi} ^3 -\nabla ^2 \tilde{\phi} ] +\tilde{\theta}, \label{nor_phiflux}
\end{equation}
and Eq. (\ref{ns}) with the Stokes approximation becomes  
\begin{equation}
   \tilde{v}_{i} = \int d\tilde{\mathbf r}^{'} \ \mathbf{T}(\tilde{\mathbf r}-\tilde{\mathbf r}^{'})_{ij}[(-\tilde{\phi} +\tilde{\phi} ^3- \nabla ^2\tilde{\phi} )\nabla \tilde{\phi} +\tilde{\zeta}_{j}, 
\end{equation}
where $\mathbf{T}_{ij}$ is the Oseen tensor given by  
\begin{equation}
     \mathbf{T}(\tilde{\mathbf{r}}) _{ij}= \frac{\mathcal{A}}{8\pi \tilde{r}}(\frac{\delta _{ij}}{\tilde r} + \frac{\tilde{r}_{i}\tilde{r}_{j}}{\tilde{r}^3}). 
\end{equation}
$\tilde{\theta}$ and $\tilde{\mathbf{\zeta}}$ are scaled thermal noises which satisfy the following fluctuation-dissipation relations: 
\begin{equation}
   \langle \tilde{\theta} (\tilde{\mathbf r},\tilde{t})\tilde{\theta} (\tilde{\mathbf r}^{'},\tilde{t}^{\ '}) \rangle = -\mathcal{B}\nabla ^2\delta (\tilde{\mathbf r}-\tilde{\mathbf r}^{'})\delta (\tilde{t}-\tilde{t}^{\ '}), 
\end{equation}
and
\begin{equation}
   \langle \tilde{\zeta} _{i}(\tilde{\mathbf r},\tilde{t}) \tilde{\zeta}_{j}(\tilde{\mathbf r}^{'},\tilde{t}^{\ '}) \rangle = -(\mathcal{B}/\mathcal{A}) \delta _{ij}
   \nabla ^2\delta (\tilde{\mathbf r}-\tilde{\mathbf r}^{'})\delta (\tilde{t}-\tilde{t}^{\ '}). 
\end{equation}
Here we note that $\sigma/(k_{\rm B}T) = (2^{3/2}/3) |\gamma| \xi \phi _{\rm e}^2$, which is the interface tension 
in an equilibrium two-liquid coexistence state.

\vspace{0.3cm}

\noindent
{\bf Simulations without thermal noise.}
We performed simulations without thermal noise to show that droplet coarsening takes place even without thermal noise ({\it i.e.}, $\mathcal{B}=0$). 
We note that the Brownain coagulation mechanism fully relies on the presence of thermal force noise.
Simulations without thermal noise are also useful to see the spatial distribution of physical quantities such as the chemical potential and the fluid velocity 
without suffering from thermal noise.  
For these simulations, we introduce Gaussian random noise into $\tilde{\phi}$ with an intensity $\delta \tilde{\phi}= 0.01$ around the averaged composition $\tilde{\phi}_{0} = -0.5$ 
({\it i.e.}, $\Phi=0.25$) as an initial condition only at $\tilde{t}=0$, switch off noise for $\tilde{t}>0$, and follow the pattern evolution process.

\vspace{0.3cm}

\noindent
{\bf Numerical scheme to solve the above set of equations properly with both composition and force noises and the details of simulations.}
Here we explain how thermal composition and velocity fluctuations can be incorporated in our simulations. 
Although we have written Eqs. (7) and (8) as two separate equations, they can be unified into only one kinetic equation in the Stokes regime, 
since the velocity field is set solely by the composition field. Substituting Eq. (8) into Eq. (7), the dynamic equation of composition is 
given by \cite{onuki,koga1993late}
\be
\frac{\partial \tilde{\phi} ({\bf \tilde{r}},\tilde{t})}{\partial \tilde{t}} = \int d{\bf \tilde{r}^{'}}L({\bf \tilde{r}},{\bf \tilde{r}^{'}})\frac{\delta \mathcal{H} }{\delta \tilde{\phi} ({\bf \tilde{r}^{'}})} + \theta ^{R}({\bf \tilde{r}},\tilde{t}), 
\ee
where the kinetic coefficient is given as 
\begin{equation}
L({\bf \tilde{r}},{\bf \tilde{r}^{'}}) = \nabla ^2 \delta ({\bf \tilde{r}}-{\bf \tilde{r}^{'}})+ \nabla \tilde{\phi} ({\bf \tilde{r}})\cdot {\bf T}({\bf \tilde{r}}-{\bf \tilde{r}^{'}}) \cdot \nabla \tilde{\phi} ({\bf \tilde{r}^{'}}) \nonumber
\end{equation}
and the random noise term $\theta ^{R} ({\bf r})= -{\bf v}^{R}({\bf \tilde{r}})\cdot \nabla \tilde{\phi} +\tilde{\theta}$ satisfies the following fluctuation-dissipation relation
\be
   \langle \theta ^R({\bf \tilde{r}},\tilde{t})\theta ^{R}({\bf \tilde{r}^{'}},\tilde{t}^{'}) \rangle = 2L({\bf \tilde{r}},{\bf \tilde{r}^{'}})\delta (\tilde{t} -\tilde{t}^{'}). 
\ee
Here the random velocity noise ${\bf v}^{R}$ is given by ${\bf v}^{R}({\bf \tilde{r}},{\bf \tilde{t}}) = \int d{\bf \tilde{r}^{'}}{\bf T}({\bf \tilde{r}}-{\bf \tilde{r}^{'}}) \tilde{\mbox{\boldmath$\zeta$}}({\bf \tilde{r}})$.
Since the kinetic coefficient of the above equation depends upon the composition field $\tilde{\phi}$, the noise term is multiplicative and should be treated via the Stratonovich interpretation \cite{Platen}. Thus we adopted a Stratonovich scheme with semi-implicit terms developed by Camley and Brown \cite{camley2010dynamic} under a periodic boundary condition 
with a staggered grid. This scheme was originally conceived to study the phase separation process in a two-dimensional membrane.  
We modify their approach by using the Oseen tensor appropriate for a three-dimensional (3D) Newtonian fluid in the Stokes regime.
Here it is worth noting that 3D simulations of model H with thermal noises were also performed by a Lattice-Boltzmann method 
\cite{thampi2011lattice}.  
The system size was set to 256$^3$ lattices, which is large enough to study the late stage of phase separation. We set the grid size to be $\Delta x = \Delta y = \Delta z = 1$. 
The volume fraction of the minority phase $\Phi$ is set to 0.25 and the composition is set to spatially homogeneous at $\tilde{t}=0$. 

\vspace{0.3cm}
\noindent
{\bf Simulations of the motion of a pair of droplets.}  
We place two droplets whose interfacial profile is described by a step function at $\tilde{t}=0$ and then follow the process of equilibration by solving the model H equations. 
In this simulation, the fluidity parameter $\mathcal{A}$ is 50 and the thermal fluctuations are not included ({\it i.e.}, $\mathcal{B}=0$). 
Here the large fluidity is used to access a large separation between the interfacial thickness 
and the droplet size within a limited simulation time as well as to see hydrodynamics effects clearly. 
Note that the process is much slower without thermal noises. The simulation box size is 192$^3$. 

\vspace{0.3cm}

\noindent
{\bf Analysis.}
We analyse the sizes and positions of droplets by counting the number of lattice points belonging to each droplet and calculating the positions of the centre of mass of the lattice points after binarization of the composition field. We also obtain the velocity of a droplet by calculating the average of velocity over lattice points inside the droplet. 
The structure factor $S(\tilde{k})$ is obtained by spherically averaging the power spectrum of the Fourier transformation of $\tilde{\phi}(\tilde{\mathbf{r}})$. 
The characteristic wavenumber $\tilde{k}_1$ is calculated as $\tilde{k}_1=\int d\tilde{k}\ \tilde{k} S(\tilde{k})/\int d\tilde{k}\ S(\tilde{k})$.

%\bibliographystyle{naturemag4}
%\bibliography{dropletPS}

\vspace{0.7cm}

\noindent
{\bf Acknowledgements} 
We thank John Russo and Taiki Yanagishima for critical reading of our manuscript. 
This study was partly
supported by Grants-in-Aid for Scientific Research (S) and Specially Promoted Research
from the Japan Society for the Promotion of Science (JSPS), the Aihara Project, the
FIRST program from JSPS, initiated by the Council for Science and Technology Policy
(CSTP). The computation in this work has partly been done using the facilities of the Supercomputer Center, 
the Institute for Solid State Physics, the University of Tokyo.

\end{document}